\documentclass[prl,nopacs,superscriptaddress,amsfonts,amsmath,floatfix,twocolumn]{revtex4}

\usepackage{graphicx}
\usepackage{float}
\usepackage[normalem]{ulem}
\usepackage{color}
\usepackage[]{units}
\usepackage{braket}
\usepackage{graphicx}
\usepackage{stmaryrd}
\usepackage{graphicx}
\usepackage[normalem]{ulem}
\usepackage{color}
\usepackage{dsfont}
\usepackage{units}
\usepackage{algorithm}
\usepackage{algpseudocode}
\usepackage{pseudocode}
\usepackage{dcolumn}
\usepackage{bm}
\usepackage{xcolor}
\usepackage{adjustbox}
\usepackage{braket}
\usepackage{bbm}
\usepackage{float}
\usepackage{epstopdf}

\newcommand{\add}[1]{\textcolor{black}{#1}}
\newcommand{\NV}{\mbox{\it NV}}
\newcommand{\RWA}{\mbox{RWA}}
\newcommand{\NM}{\mbox{NM}}

\newcommand{\affA}{Institute for Quantum Optics and Center for Integrated Quantum Science and Technology, Universit\"at Ulm, D-89081 Ulm, Germany}
\newcommand{\affB}{Institute for Complex Quantum Systems and Center for Integrated Quantum Science and Technology, Universit\"at Ulm, D-89081 Ulm, Germany}
\newcommand{\affC}{Theoretische Physik, Universit\"at des Saarlandes, D-66123 Saarbr\"ucken, Germany}

\begin{document}

\title{Autonomous Calibration of Single Spin Qubit Operations}
\author{Florian Frank}\affiliation{\affA}
\author{Thomas Unden}\affiliation{\affA}
\author{Jonathan Zoller}\affiliation{\affB}
\author{Ressa S. Said}\affiliation{\affB}
\author{Tommaso Calarco}\affiliation{\affB}
\author{Simone Montangero}\affiliation{\affB}\affiliation{\affC}
\author{Boris Naydenov}\affiliation{\affA}
\author{Fedor Jelezko}\affiliation{\affA}

\maketitle

{\bf
Fully autonomous precise control of qubits is crucial for quantum information processing, quantum communication, and quantum sensing applications. It requires minimal human intervention on the ability to model, to predict and to anticipate the quantum dynamics~\cite{Mavadia2017,Wang2017}, as well as to precisely control and calibrate single qubit operations. Here, we demonstrate single qubit autonomous calibrations via closed-loop optimisations of electron spin quantum operations in diamond. The operations are examined by quantum state and process tomographic measurements at room temperature, and their performances against systematic errors are iteratively rectified by an optimal pulse engineering algorithm. We achieve an autonomous calibrated fidelity up to 1.00~on a time scale of minutes for a spin population inversion and up to~0.98~on a time scale of hours for a Hadamard gate within the experimental error of 2\%. These results manifest a full potential for versatile quantum nanotechnologies. 
}

\begin{figure}[htbp]
\centering
\includegraphics[scale=0.42]{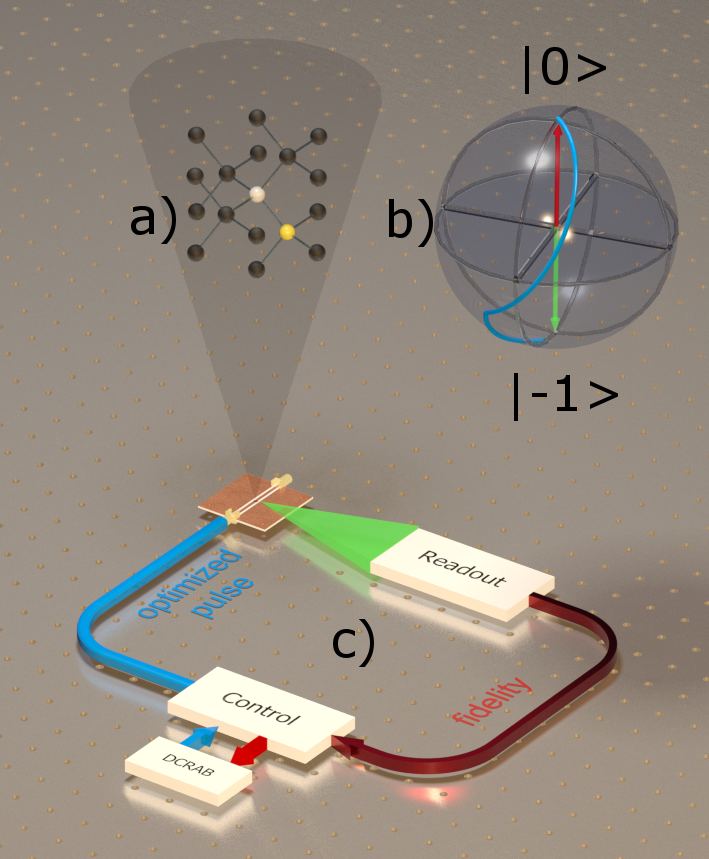}
\caption{Schematics of the experiment. The closed-loop optimisation of electron spin qubit operations interfaced to a nitrogen-vacancy (\NV) centre in diamond (nitrogen atom in yellow) (a) is  performed on a homebuilt confocal microscope (c). The spin state is initialised and readout optically, and microwave pulses are applied to manipulate the state and can be used to create gates. To start an optimisation process, a guess pulse is applied to the sample through a microwave antenna and the figure of merit is evaluated by state tomography on the spin state. This fidelity estimate is then fed to an DCRAB based algorithm and a new test-pulse is generated. The spin trajectory (blue arrow) corresponding to an optimised pulse is shown on the Bloch sphere (b). As sketched, these steps are iteratively repeated until a previously definded fidelity is reached.
}
\label{fig1}
\end{figure}
Our spin qubit implementation is a single nitrogen-vacancy~(\NV) colour centre in diamond. It provides a suitable platform for a precise qubit manipulation to be realised~\cite{Doherty2013,Wrachtrup2006}. Its remarkable features, such as optical initialisation and readout, and the ability to be manipulated by microwave fields at room temperature, make this physical system extremely attractive for many quantum technologies~\cite{Hemmer2009}. We have witnessed a vast array of demonstrations of the~\NV~centres showing a~great potential for future technologies, ranging from sub pico-Tesla magnetometry~\cite{Wolf2015}, electric field and temperature sensing~\cite{Dolde2011,Neumann2013}, to probing molecular dynamics~\cite{Staudacher2015}, and single-cell magnetic imaging~\cite{Glenn2015}. Furthermore, intertwinements between quantum information and metrology using~\NV~centre based systems yield novel and effective techniques towards the realisation of high-performance technologies, e.g. applying quantum error 
correction~\cite{Unden2016} and phase estimation~\cite{Bonato2016} to improve magnetic field sensitivity. One way to reach such technology is to apply the closed-loop optimisation method for auto-calibrating the controls required to drive the system in the presence of experimental limitations and noise. Closed-loop optimal control has been already applied to quantum information processing~\cite{Rosi2013,Brif2010}. However, to date no realisation of such autonomous calibration in room-temperature solids has been reported. Here, we apply a technique derived from optimal control theory, namely the dressed chopped random basis~(DCRAB) algorithm~\cite{Doria2011,Rach2015}, to perform real-time closed-loop optimisations of two fundamental single qubit operations, a~\mbox{spin-$\nicefrac{1}{2}$} population inversion and a Hadamard gate, against frequency detuning. The~algorithm is adapted for use in the~\NV~centre based experiment and directly embedded in the experimental apparatus, allowing the autonomous spin qubit calibrations to be fully performed as illustrated by Fig.~\ref{fig1}.

\begin{figure*}[tp]
\centering
\includegraphics[width=0.93\textwidth]{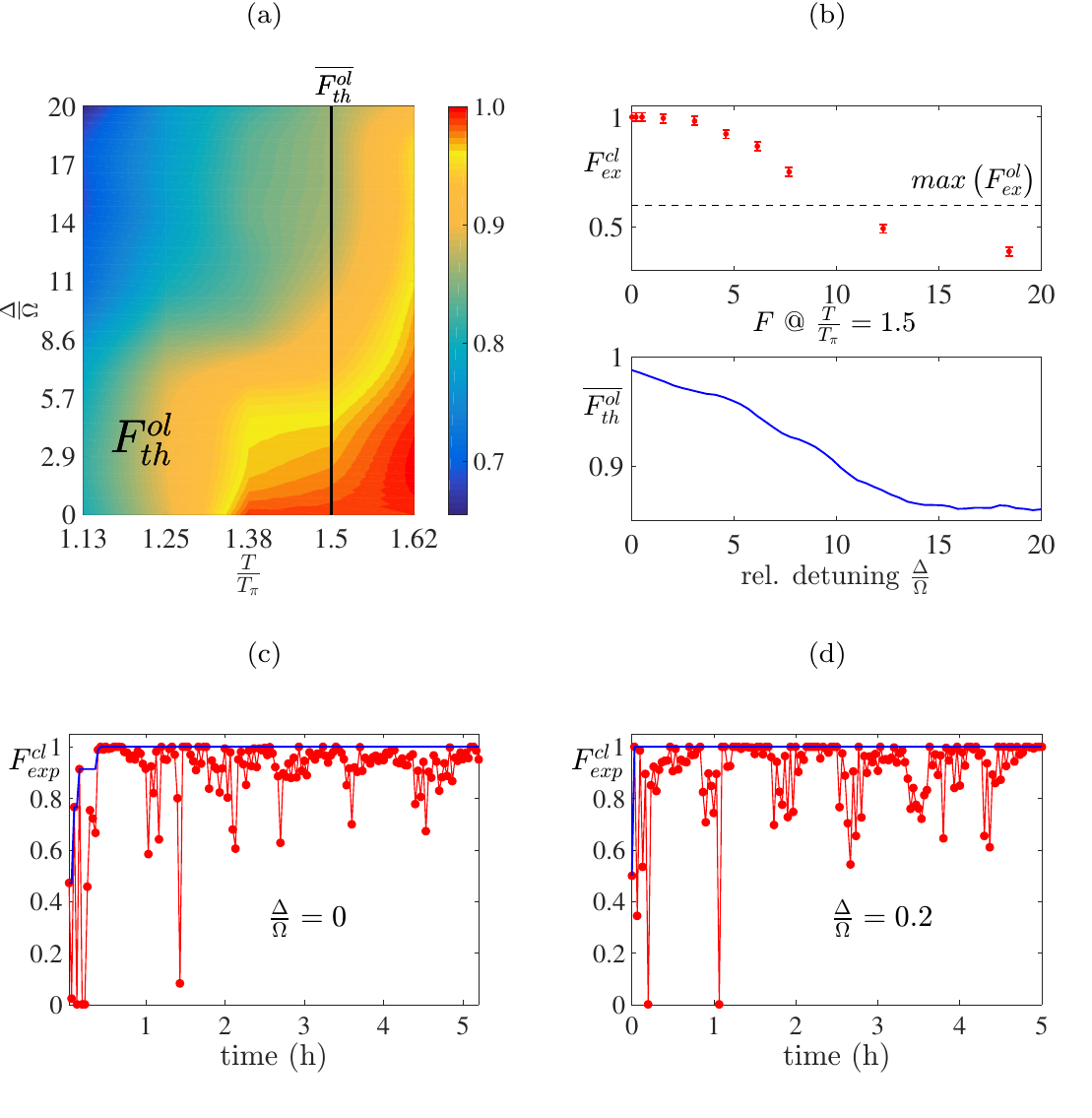}
\caption{ Optimal state transfer in presence of frequency detuning and limited time resources. (a) shows the 
resulting fidelity ($F_{ol}^{th}$) from the parameter study of the optimized state transfer as a function of relative process time $\frac{T}{T_{\pi}}$ and rel. detuning $\frac{\Delta}{\Omega}$. $T$ is the time duration of the applied pulse and $T_{\pi}$ is 
the time necessary for a population inversion when a constant Rabi frequency of $\Omega$ is applied. $\Delta$ is the detuning of the 
microwave frequency. In (b) one cross-section ($\frac{T}{T_\pi}=1.5$) of the results from (a) is compared with the results of the closed-loop experimental optimisation ($F_{cl}^{ex}$) (red points). 
Additionally, the level of best achieved fidelities from experimental evaluation of open-loop pulses is indicated (by 
black dashed line). (c) shows the evaluated fidelities ($F_{cl}^{ex}$) when the microwave pulse is applied on 
resonance and in (d) when a relative detuning of $0.2$ is applied. The blue lines show the last, highest fidelity achieved.
}\label{fig2}
\end{figure*}
\paragraph{Closed-loop optimisation\add{.}}  
Optimal control methods have been applied to several quantum information processing tasks 
with~\NV~centres~\cite{Dolde2014,Waldherr2014,Wang2015,Scheuer2014,Noebauer2015}, affirming their 
necessity and significance for quantum technology. However, the previously reported 
experiments~\cite{Dolde2014,Waldherr2014,Wang2015,Scheuer2014,Noebauer2015}, utilise {\it open-loop 
optimisation} techniques where the optimisation is performed before the actual experiment~by separate computer 
simulations. The technique requires system--environment coupling information as detailed as possible to provide a 
robust solution. In contrast, the closed-loop technique requires no explicit system--environment information. Hence, it~is utmost practical for the realisation of versatile quantum devices. One significant feature of the DCRAB algorithm is that it makes such closed-loop optimisation viable since the only quantity required from the experiment is a single figure of merit (e.g.~state or gate fidelity). No further information, such as a gradient or a Hessian, is necessary. Moreover, recent theoretical work ~\cite{Lloyd2014,Caneva2014}, points out that the relevant number of degrees of freedom in the control is rather small for few qubit systems. A reasonable number of degrees of freedom can be addressed through a suitable 
parametrisation~\cite{Viola1998,OHara2001,Wigley2016}. The DCRAB algorithm makes use of this foundation. It shapes high 
accuracy pulses with few iterations (or {\it superiteration} that is required for avoiding local optimisation traps), and maintains the  robustness against noise and errors potentially occurred at any 
stage of experiments~\cite{Rach2015}. We provide detailed discussions of the algorithm and its implementations in Sec.~A 
in the supplementary material.

\paragraph{Experimental setup\add{.}} The two-level quantum system considered in this work is built by the spin 
states~$\ket{m_s = 0}$ and $\ket{m_s = -1}$ of the ground electronic state~of an~\NV. Electron spin initialisation and readout are performed on a home built confocal microscopy setup at room temperature. To perform quantum operations on the~\NV~spin with high fidelity, the microwave field source is controlled by an arbitrary waveform generator~(AWG, Keysight M8195A), with a timing resolution of 65~GS/s. In combination, we used a 50~W amplifier with a frequency bandwidth of about 4~GHz. The microwave field was created with a copper wire close to the~\NV. By controlling the amplitude and the phase of the microwave, we are able to rotate the system spin around arbitrary axis on the Bloch sphere. The time dependent control Hamiltonian is given by
\begin{equation}
H_c = 2\pi \Omega\left(X(t) \hat{S_x}+Y(t) \hat{S_y}\right),
\end{equation}
while the system Hamiltonian in the rotating frame is 
$
H_0 = 2\pi \Delta \hat{S_z},
$
where $\hat{S_x}$, $\hat{S_y}$ and $\hat{S_z}$ are the spin operator of a two-level system. The functions $X(t)$ and $Y(t)$ define the corresponding microwave pulse, where $\Omega$ is the Rabi frequency and $\Delta$ is the microwave frequency detuning from energy splitting of our two level system. In our experiments the following conditions were fulfilled,
\begin{eqnarray}
X(t)+Y(t) \in \left[-1~1\right], \,
\Omega \in \left[0~10\right] \textnormal{(in MHz).}
\end{eqnarray}
For further proceeding, we define the relative detuning $\frac{\Delta}{\Omega}$ and the relative process time $\frac{T}{T_{\pi}}$, with $T_{\pi}=\frac{1}{2\Omega}$ and the overall process time is $T$.

\begin{figure*}[tp]
\includegraphics[width=0.93\textwidth]{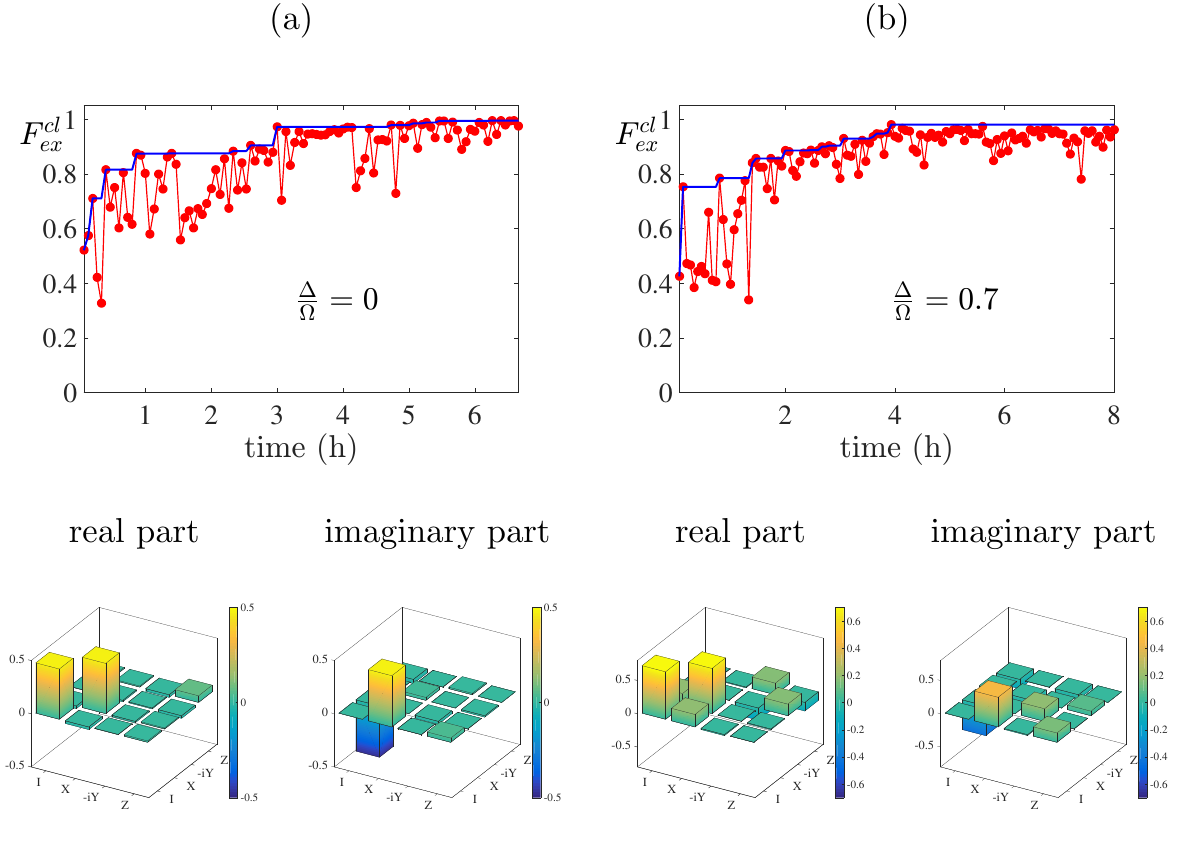}
\caption{Optimisation of a Hadamard gate when the control pulse is applied on resonance with the \NV~ transition (a) and when a detuning on the order of the Rabi frequency is artificially applied (b). For each scenario we show the measured fidelity $F_{ex}^{cl}$ with increasing number of function evaluations and the experimentally obtained process matrix for the real (left) and the imaginary part (right) in case of the highest evaluated fidelity.}\label{fig3}
\end{figure*}

\paragraph{High fidelity population inversion via closed-loop control\add{.}}
In this experiment we search for an optimal microwave pulse to transfer the NV spin state $\ket{m_s=0}$ to 
$\ket{m_s=-1}$ with high fidelity. Therefore, we performed a state tomography after applying a parametrised microwave 
pulse. Subsequent measurement of the state transfer fidelity is then used as a 
figure of merit. To estimate the particular fidelity $F_{ex}^{cl}$ we utilised quantum state tomography (see Sec. B and Sec. C the 
supplementary information). For convention, the subscripts of the fidelity $F_{i}^{j}$ indicate either experimental $(ex)$ or theoretical $(th)$ data. The superscripts $(cl)$ and $(ol)$ refer to closed-loop and open-loop control, respectively.
In Fig.~\ref{fig2} we show the results of our state transfer optimisation. In panel (a), we first identify via an 
open-loop simulation the expected performance of the employed optimisation method in presence of limited 
transfer time and static detuning. For relative process times $T/T_{\pi}$ exceeding $\sim 1.5$, our simulations identify 
robust solutions even up to a relative detuning of about $\frac{\Delta}{\Omega} = 10$. Our experimental results (see panel b) achieved via closed loop optimisation support these results for small 
detuning, when $T/T_{\pi}=1.5$. Each optimisation, performed for a certain gate time and a certain detuning, bases on a DCRAB 
algorithm with 6 superiterations. Exemplarily, we show in Fig.~\ref{fig2}~(c,d) the full closed-loop optimisation 
process in the case of no detuning, and when a small detuning is applied. 
The blue curve shows the currently best found solution, while the red line is an internal algorithmic figure of merit quantity (for 
details see Supp. Mat. Section A).
The necessary time for achieving optimal 
fidelities depends on the accuracy of the tomography measurements ($F_{ex}^{cl}$). We achieve the maximal fidelity of 1.00 with an accuracy of 
$10^{-2}$ on a reliable timescale of about $2000$ seconds. It is interesting to note, that any optimal pulse of the 
open-loop simulation was not able to beat the results of our closed-loop strategy in the case of moderate detuning. The 
best fidelity achieved via open-loop techniques ($F_{th}^{ol}$) is marked in Fig.~\ref{fig2}~(b) and is only on the order of $0.6$.   
 
\paragraph{\color{black} Hadamard gate fidelity auto--calibration\add{.}}
A $\frac{\pi}{2}$-gate is the basic block of generating coherent quantum processes like quantum metrology and quantum computing. To show the capabiliites of our concept, we subsequently optimized the quantum gate
\begin{align}
G = \frac{1}{\sqrt{2}}
\begin{pmatrix}
1 & -i \\
-i & 1 
\end{pmatrix},
\end{align}
which is a Hadamard gate up to a global phase. In accordance to the first experiment, we first performed the experiment with no detuning and second, when a detuning of $8.125$~MHz (relative detuning $\frac{\Delta}{\Omega}=0.7$) was applied. The results are shown in Fig.\ref{fig3}. To quantify the perfomance of the evaluated microwave pulse with respect to the defined quantum gate $G$, quantum process tomography was used (for details see Supp. Mat. Section D and E). We observe after $99$ evaluations a fidelity of $0.99\pm 0.01$ and after $58$ evaluations a fidelity of $0.98\pm0.02$ when the static microwave detuning was applied. The initial fidelity of the guess pulse was in both cases about $0.50$. Compared to the previous experiment, the time needed for optimal results is about $4$ times longer due to additional measurements, which are necessary for complete process tomography.

\paragraph{Discussion and Outlook\add{.}}  
Our experimental results demonstrate that the closed-loop feedback control overcomes static and unknown system errors to achieve the high fidelity autonomous calibration of single quantum gates that is necessary for future quantum technologies with room temperature solids. Our approach of the closed-loop optimisation uses minimal control resources and experimental knowledge that are accessible for users. The total time, required for autonomous calibration, is mainly determined by the duration of quantum tomography measurement and not by the optimisation algorithm. Hence, a significant speedup in the total time of calibration and its fidelity precision may potentially be achieved by employing fast and simplified tomography methods, for instance randomised benchmarking~\cite{Knill2008}. In addition to the autonomous calibration, our demonstrated closed-loop optimisation features stabilisation mechanism against experimental drifts, for instance due to a misalignment of the permanent magnet used to lift the ground state degeneracy. Our procedure presented in this letter is not limited for application to single-qubit operations only. Further experimental implementations towards multi-qubit gate autonomous calibrations are in principle feasible using our closed-loop optimisation method.

\vspace{0.2cm}
\paragraph{Acknowledgements\add{.}}
We acknowledge financial supports from European Union FET Projects DIADEMS, RYSQ, and SIQS, German Research Foundation (DFG) through the SFB/TRR21 projects, German Federal Ministry of Education and Research (BMBF) via the Q.Com projects, and the Center for Integrated Quantum Science and Technology~(IQST). We thank S. Zaiser, P. Neumann, and J. Wrachtrup in Stuttgart University for their valuable support and discussions. S.M. gratefully acknowledges the support of the DFG via a Heisenberg fellowship.

\vspace{0.2cm}
\paragraph{Author contributions\add{.}}  
F.F., T.U., and J.Z. contributed equally to this work. F.F., T.U., and B.N. carried out the experiment and analysed the data. J.Z. and R.S.S. provided theoretical supports and performed numerical simulations and analyses. F.F., T.U., J.Z., and R.S.S wrote the manuscript with feedback from all authors. T.C., S.M., B.N. and F.J. supervised and managed the project.

\newpage

\section{Supplementary Material}

\subsection{A. Dressed CRAB Algorithms}
The chopped random-basis algorithm~(CRAB) is originally proposed to provide one simple yet~powerful numerical optimisation method for controlling complex quantum many-body dynamics, where the gradient of~the figure of~merit to be optimised is hard to obtain and the controlling apparatus is realistically bandwidth-limited~\mbox{\cite{SM_Doria2011,SM_Caneva2011}}. Recently, an~extended version of~the~algorithm, namely the dressed~CRAB~\mbox{(DCRAB)}, has been developed to tackle possible local minima or~false traps in the optimisation landscape~\cite{SM_Rach2015}. The standard version of CRAB algorithm has experimentally been implemented and adapted in~various physical systems, for instance in nitrogen-vacancy~(\NV) centre in diamond for precise spin controls beyond rotating-wave-approximation~(\RWA)~\cite{SM_Scheuer2014}, in ultra-cold atoms in optical lattice for controlling quantum phase-transition from a superfluid to a~Mott insulator~\cite{SM_Rosi2013}, in a non-classical state of Bose-Einstein condensate~(BEC)~\cite{SM_Frank2014}, in superconducting qubits~\cite{SM_Brouzos2015,SM_Alon2016}, and in a hybrid system of~BECs and~cold atoms in optical lattice~\cite{SM_Frank2016}.~A~theoretical relation to quantum speed limit in many-body quantum system has also been studied~\cite{SM_Caneva2009}.  
\begin{figure}[h]
\vspace{0.2cm}
\centering
\includegraphics[scale=0.85]{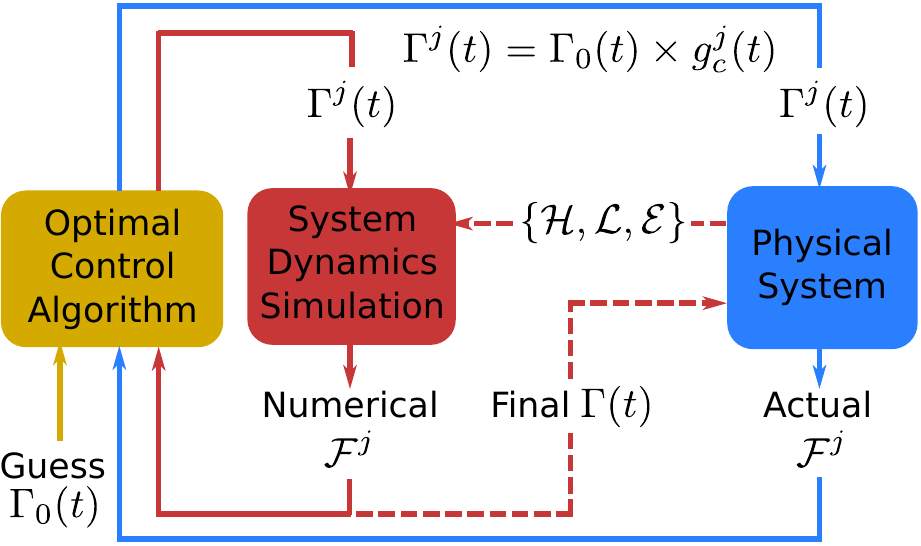}
\vspace{2pt}
\caption{\label{open_vs_close}(Color online) 
{\it Open-loop} and~{\it closed-loop} optimisation methods. An~optimised or a~guess pulse~$\Gamma_0\left(t\right)$, is fed into an optimal control algorithm (as gold, and in our work this is DCRAB algorithm). For each update (denoted by~$j$, and can be either {\it function evaluation} or {\it iteration}), the~{\it open-loop} method (red) uses a system dynamics simulator, e.g. a~personal computer~(PC) or a~cluster of~PCs, to numerically provide a single optimised figure of~merit~$\mathcal{F}^j$. The simulator requires a set of information obtained indirectly from the actual physical system: a~controllable Hamiltonian~$\mathcal{H}$, a~system-environment interaction model~$\mathcal{L}$, and a~description of~noises or~errors~$\mathcal{E}$. If the numerical~$\mathcal{F}^j$ reaches a certain target value, the process is terminated and the final optimised pulse~$\Gamma\left(t\right)$, is passed to the actual physical process. In case the set of information from the actual physical system is not available or can only be partially obtained, one can resort to the~{\it closed-loop} optimisation method~(blue). The actual physical system acts as a simulator which outputs the actual~$\mathcal{F}^j$. If the target~$\mathcal{F}^j$ is satisfied, the desired process is directly started. In principle, both methods can be used complementarily as a~{\it hybrid} method.
}
\end{figure}

Here, we elaborate the DCRAB algorithms adapted for our current work on the real-time {\it close-loop} optimisation of \NV~electron spin control. To begin with, we provide in Fig.~\ref{open_vs_close} a~sketch of the {\it open-loop} and~{\it closed-loop} optimisation methods.{\color{black} The~{\it closed-loop} method offers users the ability to obtain an optimised control pulse for manipulating the system without having a full access to the complete information about it.}

Given an unoptimised microwave pulse as a guess pulse~$\Gamma_0\left(t\right)$, that controls the \NV~electron spin ground state, the DCRAB algorithm iteratively finds an updating pulse~$g_c^{j,k}(t)$, which gives an optimal pulse~$\Gamma(t)^{j,k}=\Gamma_0(t)\times g_c^{j,k}(t)$, such that a~certain figure of~merit~$\mathcal{F}^{j,k}=\mathcal{F}^{j,k} [\Gamma(t)^{j,k}]$ is~maximised (or~minimised, if~one uses~$1-\mathcal{F}^{j,k}$), for a predetermined and fixed duration of time~$t=T$. Here, we use $j$ to denote the update which can be either {\it function evaluation} or {\it iteration},~\footnote{We distinguish between these quantities since one iteration may require multiple function evaluations.} and $k$ to represent~{\it super-iteration}. In a direct search method introduced later in this section, the figure of~merit does not necessarily improve over a number of increasing function evaluations, nevertheless, it is not decreasing. The typical examples of~figure~of merit are {\it state-to-state} or state fidelity, quantum gate fidelity, and entropy of entanglement~\cite{SM_Caneva2011}. Following Ref.~\cite{SM_Rach2015}, one may expand the updating pulse~$g_c^{j,k}\left(t\right)$, as 
\begin{eqnarray}
g_c^{j,k}\left(t\right) = \frac{1}{\lambda\left(t\right)} \sum_{n=0}^N a_n^{j,k} \sin \omega_n^{j,k} t + b_n^{j,k} \cos \omega_n^{j,k} t.
\label{dcrabupdate}
\end{eqnarray}
where $N$~is commonly termed as a number of DCRAB {\it frequency components}, and $\omega_n^{j,k}=2\pi(n+r)/T$~is {\it a randomised frequency} due to a~random number~\mbox{$|r|<0.5$}~\cite{SM_Scheuer2014}. The function~$\lambda(t)$ is~a~predetermined function limiting the update pulse to be zero at~$t=0$ and~$t=T$. It allows for a smooth ramping at the beginning and at the end of the pulse. This limiting function is flexibly chosen and its parameters depend on some experimental factors, e.g.~the resolution of the controlling apparatus~(see Ref.~\cite{SM_Scheuer2014} for further details). For each~{\it super-iteration}~(after a certain number of function evaluation~$j=J$), DCRAB performs re-optimisations by transforming the updating pulse~\mbox{$g_c^{J,k-1} \mapsto g_c^{J,k-1} + g_c^{J,k}$}, where the parameters~{$\{a_n^{J,k},b_n^{J,k}\}$} remain the same, while the randomised frequencies are updated~{$\omega_n^{J,k-1} \mapsto \omega_n^{J,k}$}. 
From Eq.~\ref{dcrabupdate}, one finds that the optimisation problem is reduced to a search problem: 
{``Find a~set of~parameters~$\{a_n^{j,k},b_n^{j,k}\}$, which maximises~$\mathcal{F}^{j,k}$, given that the DCRAB frequency components does not change at every function evaluations or iterations (i.e.~$\{a_n^{j,k},b_n^{j,k},\omega_n^{j,k}\}\mapsto\{a_n^{j+1,k},b_n^{j+1,k},\omega_n^{j,k}\}$)."} 

In principle, any direct search algorithm can be employed to solve this problem, such as Nelder-Mead or simplex algorithm~\cite{SM_Nelder1965}, Powell algorithm~\cite{SM_Powell1964}, and pattern search algorithm~\cite{SM_Audet2002}. Here, we employ the Nelder-Mead~(\NM) algorithm since it has shown its strength in optimizing various unconstrained minimization problems~\cite{SM_Lagarias1998,SM_Lewis2000}, and has been implemented and efficiently tested in many code libraries, e.g.~PYTHON~\cite{SM_optimize} or~MATLAB~\cite{SM_fminsearch}. We encode the DCRAB method with the~\NM~algorithm in PYTHON, and directly embed it on a PC interfaced with the microwave control apparatuses, for instance an~arbitrary wave generator~(AWG)~\cite{SM_Frank2015}.

\begin{figure}[tp]
\vspace{0.2cm}
\centering
\includegraphics[scale=0.6]{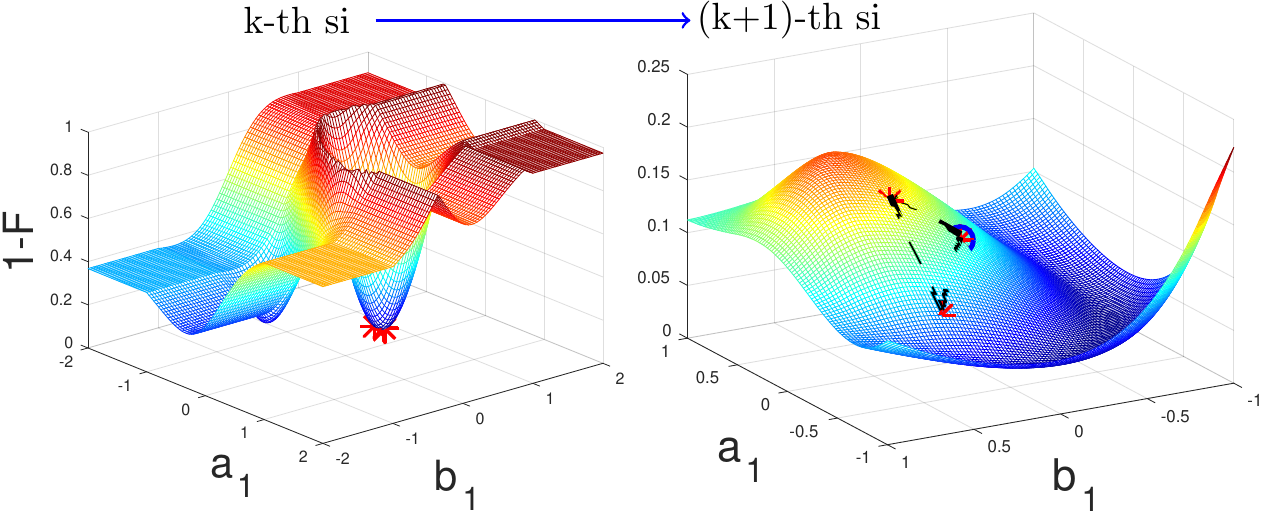}
\vspace{2pt}
\caption{\label{fig_land}(Color online) 
Control landscape of the single qubit gate synthesis at two stages of optimisation. The Nelder-Mead search at the k-th {\it super-iteration}~(si) goes towards the marked local minimum or trap (the left pane). Starting at that trap where the bases of the pulse information has been gained (the blue circle on the right pane), the subsequent search with the basis changes unlocks the local trap allowing the optimisation to proceed further downhill. The dashed black triangle shows the possible span of the Nelder-Mead simplex.
}
\end{figure}

A numerical evidence of the DCRAB €™capabilities is depicted in Fig.~\ref{fig_land} which shows two instances of a control landscape. Control landscapes display the quality (e.g. fidelity) of an optimisation as a function of the control variables~\cite{SM_Bayer2008,SM_Roslund2009,SM_Roslund2009a}. Here, the control variables are the amplitude and phase of a sinusoidal curve. Due to the limited function space to be explored at a time, the algorithm might get trapped at a sup-optimal point in the landscape. Further sub-optimal points may be induced by restrictions unavoidably imposed by the lab environment such as limited bandwidth, control power limitations, and noise occurring at various stages throughout the experiment. By moving on to the next super-iteration and hence doing a basis change (see arrow in the figure), the trap is unlocked with probability one allowing the algorithm to proceed and converge to a global optimum~\cite{SM_Rach2015}. 

\subsection{B. Quantum State Tomography}

In our experiments the quantum system of interest is defined by the \NV~centre's ground state spin levels $\ket{m_s=0}$ and $\ket{m_s=-1}$. An experimentally prepared density matrix $\rho$, is measured by performing two Rabi experiments each with orthogonal rotation-axis $x$ and $y$, and initial state $\rho$. The rotation axis is controlled by the phase of the resonant microwave control. In the case of the $x$-rotation~\cite{SM_Neumann1326}, after measuring the population~$P(\ket{0},t)$ in the state $\ket{m_s=0}$, we observe that
\begin{equation}
P(\ket{0},t) = \frac{d+a}{2}+\frac{d-a}{2}\cos(2\pi\omega t)-c \sin(2\pi\omega t).
\end{equation}
In the case of the $y$-rotation, we have 
\begin{equation}
P(\ket{0},t) = \frac{d+a}{2}+\frac{d-a}{2}\cos(2\pi\omega t)+b \sin(2\pi\omega t).
\end{equation}
Here, the applied microwave duration is denoted by~$t$, and the initial density matrix is given by
\begin{equation}
\rho= \begin{pmatrix} \rho_{\ket{-1}\bra{-1}} & \rho_{\ket{-1}\bra{0}}  \\ \rho_{\ket{0}\bra{-1}} & \rho_{\ket{0},\bra{0}} \end{pmatrix}=\begin{pmatrix} a & b-ic  \\ b+ic & d \end{pmatrix}.
\end{equation}
Therefore, we identify the entries of the initial density matrix $\rho$ by fitting the experimentally observed Rabi oscillations to the described model. To guarantee an allowed density matrix, we use quadratic maximum likelihood estimation to extract the density matrix, which is parameterised accordingly by
\begin{equation}
\rho(\xi,\nu)=e^{-i\nu S_y}e^{-i\xi S_x}\begin{pmatrix} 1 & 0  \\ 0 & 0 \end{pmatrix}e^{i\xi S_x}e^{i\nu S_y}.
\end{equation}
Here, $S_x$ and $S_y$ are the spin-$1/2$ operators. We identify the error of the estimation of a single entry of $\rho$, by 
\begin{equation}  
\sigma(\rho_{ij})=\sqrt{Var(\rho_{ij})}=\frac{1}{4}\sqrt{\sum_{i,j}\left(\rho_{ij}-\rho_{ij}(\xi,\nu)\right)^2}.
\end{equation}

\subsection{C. Fidelity Estimation: Optimising State Transfer}

The fidelity of the state transfer $\ket{m_s=0}\rightarrow\ket{m_s=-1}$ ($\ket{-1}$), is estimated by the following tasks: (i)~initialising the system in $\ket{m_s=0}$ ($\ket{0}$), (ii)~applying the gate, and (iii)~performing quantum state tomography. The fidelity is given by
\begin{equation}
F = \rho_{\ket{-1}\bra{-1}}\pm \sigma(\rho_{\ket{-1}\bra{-1}}).
\end{equation}

\subsection{D. Fidelity Estimation: Optimising a Hadamard Gate}

The fidelity of a gate is estimated by firstly initialising the system in one of the four states, $\psi_1=\ket{0}$, $\psi_2=\ket{-1}$, $\psi_3=\frac{\ket{0}-i\ket{-1}}{\sqrt{2}}$ and $\psi_4=\frac{\ket{0}+\ket{-1}}{\sqrt{2}}$. It is then followed by applying the gate, and applying the inverse of the ideal gate. Lastly, the population according to the initial state $\psi_i$, is measured by quantum state tomography. If the final state matches the initial state, the the actual gate reaches the ideal gate. After applying this process for the full set of initial states $\psi_i$, the fidelity~$F$, is estimated by
\begin{equation}
F=E(\rho_{\ket{\psi_i}\bra{\psi_i}})\pm E(\sigma(\rho_{\ket{\psi_i}\bra{\psi_i}})),
\end{equation}
where $E(\cdot)$ denotes the expectation value, and $\sigma(\cdot)$ is the standard deviation. The single qubit rotations that is necessary for the initialisation are repetitively calibrated. The corresponding fidelity is above $99.8\%$.

\subsection{E. Quantum Process Tomography}

Firstly, the system is initialised in one of the four states $\rho^0_i=\psi_i\psi_i^\dagger$. Secondly, the gate is applied, and the final state~$\rho^f_i$ is measured by quantum state tomography. We choose the following complete set of quantum operators:
\begin{align}
e_1&=\mathbbm{1},\\
e_2&=\sigma_x,\\
e_3&=-i\sigma_y,\\
e_4&=\sigma_z,
\end{align}
where $\sigma_{x/y/z}$ are the Pauli matrices. If we define
\begin{equation}
M=\begin{pmatrix} 0\cdot\mathbbm{1} & 0\cdot\mathbbm{1} & 0\cdot\mathbbm{1} & 1\cdot \mathbbm{1}\\
1\cdot\mathbbm{1} & 0\cdot\mathbbm{1} & 0\cdot\mathbbm{1} & 0\cdot\mathbbm{1} \\
0.5\cdot\mathbbm{1} & -0.5i\cdot\mathbbm{1} & 0.5i\cdot\mathbbm{1} & 0.5\cdot\mathbbm{1} \\
0.5\cdot\mathbbm{1} & -0.5\cdot\mathbbm{1} & -0.5\cdot\mathbbm{1} & 0.5\cdot\mathbbm{1}, 
\end{pmatrix}
\end{equation}
and
\begin{equation}
\beta=\begin{pmatrix} \mathbbm{1} & e_1 \\
e_1 & -\mathbbm{1} 
\end{pmatrix},
\end{equation}
the process matrix is then defined by~\cite{SM_Nielsen2011}
\begin{equation}
\chi=\beta\begin{pmatrix} M^{-1}\rho^f_1 & M^{-1}\rho^f_2 \\
M^{-1}\rho^f_3 & M^{-1}\rho^f_4
\end{pmatrix}\beta.
\end{equation}

\subsection{F. Realistic Parameters Analysis}

We perform numerical simulations on the optimised population inversion for a certain range of predetermined detuning~$\omega$, and control time~$T$, to demonstrate the robustness of the DCRAB algorithm. The simulated detuning is in the unit of Rabi frequency~$\omega_r$, and we use~$T_{\pi}$, to denote the control time of the theoretical rectangular $\pi$-pulse with no detuning. The result of the simulations is presented in Fig.~2(a) in the main manuscript. The Hamiltonian used in the simulations is
\begin{equation}
 H = \frac{2\omega}{\omega_r} \begin{pmatrix} 0 & 0 \\ 0 & -1 \end{pmatrix} + \Gamma_x(t) \begin{pmatrix} 0 & 1 
\\ 1 & 0 \end{pmatrix} +  \Gamma_y(t) \begin{pmatrix} 0 & -i 
\\ i & 0 \end{pmatrix}.
\end{equation}
The control amplitudes~$\Gamma_x(t)$ and~$\Gamma_y(t)$, are constrained by a condition of~$|\Gamma_x(t) + \Gamma_y(t)| \le 1, \; \forall t$, which corresponds to the voltage output limit of the AWG used in the experiment. 

In performing the simulations, the algorithm and its parameters~(e.g. the total number of function evaluations and  super-iterations, the update parameters, the algorithm stopping criteria, and the pulses bandwidth) are taken to be the same as those used for generating the real pulse in the closed-loop optimisation experiment. Each of the fidelities, shown~in~Fig.~2(a) in the main manuscript, are the average fidelities over 20 independent runs of initialisation and optimisation.

Finally, the set of pulses obtained from the simulation for $T=1.5\times T_{\pi}$, are directly applied to the experiment without any further closed-loop optimisation, and yields a state inversion fidelity of less than~0.6. This significant drop in fidelity is due to the experimental imperfections, the unknown system parameters, and the possible further 
constraints which are always present at any realistic setup. On the contrary, the closed-loop control optimisations take all the above-mentioned factors into account, and hence provide far better results.


\begin{thebibliography}{27}
\expandafter\ifx\csname natexlab\endcsname\relax\def\natexlab#1{#1}\fi
\expandafter\ifx\csname bibnamefont\endcsname\relax
  \def\bibnamefont#1{#1}\fi
\expandafter\ifx\csname bibfnamefont\endcsname\relax
  \def\bibfnamefont#1{#1}\fi
\expandafter\ifx\csname citenamefont\endcsname\relax
  \def\citenamefont#1{#1}\fi
\expandafter\ifx\csname url\endcsname\relax
  \def\url#1{\texttt{#1}}\fi
\expandafter\ifx\csname urlprefix\endcsname\relax\def\urlprefix{URL }\fi
\providecommand{\bibinfo}[2]{#2}
\providecommand{\eprint}[2][]{\url{#2}}

\bibitem[{\citenamefont{Mavadia et~al.}(2017)\citenamefont{Mavadia, Frey,
  Sastrawan, Dona, and Biercuk}}]{Mavadia2017}
\bibinfo{author}{\bibfnamefont{S.}~\bibnamefont{Mavadia}},
  \bibinfo{author}{\bibfnamefont{V.}~\bibnamefont{Frey}},
  \bibinfo{author}{\bibfnamefont{J.}~\bibnamefont{Sastrawan}},
  \bibinfo{author}{\bibfnamefont{S.}~\bibnamefont{Dona}}, \bibnamefont{and}
  \bibinfo{author}{\bibfnamefont{M.~J.} \bibnamefont{Biercuk}},
  \bibinfo{journal}{Nat. Commun.} \textbf{\bibinfo{volume}{8}},
  \bibinfo{pages}{14106} (\bibinfo{year}{2017}).

\bibitem[{\citenamefont{Wang et~al.}(2017)\citenamefont{Wang, Paesani,
  Santagati, Knauer, Gentile, Wiebe, Petruzzella, O{$[$}rsquor{$]$}Brien,
  Rarity, Laing et~al.}}]{Wang2017}
\bibinfo{author}{\bibfnamefont{J.}~\bibnamefont{Wang}},
  \bibinfo{author}{\bibfnamefont{S.}~\bibnamefont{Paesani}},
  \bibinfo{author}{\bibfnamefont{R.}~\bibnamefont{Santagati}},
  \bibinfo{author}{\bibfnamefont{S.}~\bibnamefont{Knauer}},
  \bibinfo{author}{\bibfnamefont{A.~A.} \bibnamefont{Gentile}},
  \bibinfo{author}{\bibfnamefont{N.}~\bibnamefont{Wiebe}},
  \bibinfo{author}{\bibfnamefont{M.}~\bibnamefont{Petruzzella}},
  \bibinfo{author}{\bibfnamefont{J.~L.} \bibnamefont{O{$[$}rsquor{$]$}Brien}},
  \bibinfo{author}{\bibfnamefont{J.~G.} \bibnamefont{Rarity}},
  \bibinfo{author}{\bibfnamefont{A.}~\bibnamefont{Laing}},
  \bibnamefont{et~al.}, \bibinfo{journal}{Nature Phys.}
  \textbf{\bibinfo{volume}{advance online publication}},
  (\bibinfo{year}{2017}).

\bibitem[{\citenamefont{Doherty et~al.}(2013)\citenamefont{Doherty, Manson,
  Delaney, Jelezko, Wrachtrup, and Hollenberg}}]{Doherty2013}
\bibinfo{author}{\bibfnamefont{M.~W.} \bibnamefont{Doherty}},
  \bibinfo{author}{\bibfnamefont{N.~B.} \bibnamefont{Manson}},
  \bibinfo{author}{\bibfnamefont{P.}~\bibnamefont{Delaney}},
  \bibinfo{author}{\bibfnamefont{F.}~\bibnamefont{Jelezko}},
  \bibinfo{author}{\bibfnamefont{J.}~\bibnamefont{Wrachtrup}},
  \bibnamefont{and} \bibinfo{author}{\bibfnamefont{L.~C.}
  \bibnamefont{Hollenberg}}, \bibinfo{journal}{Phys. Rep.}
  \textbf{\bibinfo{volume}{528}}, \bibinfo{pages}{1 } (\bibinfo{year}{2013}).

\bibitem[{\citenamefont{Wrachtrup and Jelezko}(2006)}]{Wrachtrup2006}
\bibinfo{author}{\bibfnamefont{J.}~\bibnamefont{Wrachtrup}} \bibnamefont{and}
  \bibinfo{author}{\bibfnamefont{F.}~\bibnamefont{Jelezko}},
  \bibinfo{journal}{J. Phys.: Condens. Matter} \textbf{\bibinfo{volume}{18}},
  \bibinfo{pages}{S807} (\bibinfo{year}{2006}).

\bibitem[{\citenamefont{Hemmer and Wrachtrup}(2009)}]{Hemmer2009}
\bibinfo{author}{\bibfnamefont{P.}~\bibnamefont{Hemmer}} \bibnamefont{and}
  \bibinfo{author}{\bibfnamefont{J.}~\bibnamefont{Wrachtrup}},
  \bibinfo{journal}{Science} \textbf{\bibinfo{volume}{324}},
  \bibinfo{pages}{473} (\bibinfo{year}{2009}).

\bibitem[{\citenamefont{Wolf et~al.}(2015)\citenamefont{Wolf, Neumann,
  Nakamura, Sumiya, Ohshima, Isoya, and Wrachtrup}}]{Wolf2015}
\bibinfo{author}{\bibfnamefont{T.}~\bibnamefont{Wolf}},
  \bibinfo{author}{\bibfnamefont{P.}~\bibnamefont{Neumann}},
  \bibinfo{author}{\bibfnamefont{K.}~\bibnamefont{Nakamura}},
  \bibinfo{author}{\bibfnamefont{H.}~\bibnamefont{Sumiya}},
  \bibinfo{author}{\bibfnamefont{T.}~\bibnamefont{Ohshima}},
  \bibinfo{author}{\bibfnamefont{J.}~\bibnamefont{Isoya}}, \bibnamefont{and}
  \bibinfo{author}{\bibfnamefont{J.}~\bibnamefont{Wrachtrup}},
  \bibinfo{journal}{Phys. Rev. X} \textbf{\bibinfo{volume}{5}},
  \bibinfo{pages}{041001} (\bibinfo{year}{2015}).

\bibitem[{\citenamefont{Dolde et~al.}(2011)\citenamefont{Dolde, Fedder,
  Doherty, Nobauer, Rempp, Balasubramanian, Wolf, Reinhard, Hollenberg, Jelezko
  et~al.}}]{Dolde2011}
\bibinfo{author}{\bibfnamefont{F.}~\bibnamefont{Dolde}},
  \bibinfo{author}{\bibfnamefont{H.}~\bibnamefont{Fedder}},
  \bibinfo{author}{\bibfnamefont{M.~W.} \bibnamefont{Doherty}},
  \bibinfo{author}{\bibfnamefont{T.}~\bibnamefont{Nobauer}},
  \bibinfo{author}{\bibfnamefont{F.}~\bibnamefont{Rempp}},
  \bibinfo{author}{\bibfnamefont{G.}~\bibnamefont{Balasubramanian}},
  \bibinfo{author}{\bibfnamefont{T.}~\bibnamefont{Wolf}},
  \bibinfo{author}{\bibfnamefont{F.}~\bibnamefont{Reinhard}},
  \bibinfo{author}{\bibfnamefont{L.~C.~L.} \bibnamefont{Hollenberg}},
  \bibinfo{author}{\bibfnamefont{F.}~\bibnamefont{Jelezko}},
  \bibnamefont{et~al.}, \bibinfo{journal}{Nature Phys.}
  \textbf{\bibinfo{volume}{7}}, \bibinfo{pages}{459} (\bibinfo{year}{2011}).

\bibitem[{\citenamefont{Neumann et~al.}(2013)\citenamefont{Neumann, Jakobi,
  Dolde, Burk, Reuter, Waldherr, Honert, Wolf, Brunner, Shim
  et~al.}}]{Neumann2013}
\bibinfo{author}{\bibfnamefont{P.}~\bibnamefont{Neumann}},
  \bibinfo{author}{\bibfnamefont{I.}~\bibnamefont{Jakobi}},
  \bibinfo{author}{\bibfnamefont{F.}~\bibnamefont{Dolde}},
  \bibinfo{author}{\bibfnamefont{C.}~\bibnamefont{Burk}},
  \bibinfo{author}{\bibfnamefont{R.}~\bibnamefont{Reuter}},
  \bibinfo{author}{\bibfnamefont{G.}~\bibnamefont{Waldherr}},
  \bibinfo{author}{\bibfnamefont{J.}~\bibnamefont{Honert}},
  \bibinfo{author}{\bibfnamefont{T.}~\bibnamefont{Wolf}},
  \bibinfo{author}{\bibfnamefont{A.}~\bibnamefont{Brunner}},
  \bibinfo{author}{\bibfnamefont{J.~H.} \bibnamefont{Shim}},
  \bibnamefont{et~al.}, \bibinfo{journal}{Nano Lett.}
  \textbf{\bibinfo{volume}{13}}, \bibinfo{pages}{2738} (\bibinfo{year}{2013}).

\bibitem[{\citenamefont{Staudacher et~al.}(2015)\citenamefont{Staudacher,
  Raatz, Pezzagna, Meijer, Reinhard, Meriles, and Wrachtrup}}]{Staudacher2015}
\bibinfo{author}{\bibfnamefont{T.}~\bibnamefont{Staudacher}},
  \bibinfo{author}{\bibfnamefont{N.}~\bibnamefont{Raatz}},
  \bibinfo{author}{\bibfnamefont{S.}~\bibnamefont{Pezzagna}},
  \bibinfo{author}{\bibfnamefont{J.}~\bibnamefont{Meijer}},
  \bibinfo{author}{\bibfnamefont{F.}~\bibnamefont{Reinhard}},
  \bibinfo{author}{\bibfnamefont{C.~A.} \bibnamefont{Meriles}},
  \bibnamefont{and}
  \bibinfo{author}{\bibfnamefont{J.}~\bibnamefont{Wrachtrup}},
  \bibinfo{journal}{Nat. Commun.} \textbf{\bibinfo{volume}{6}},
  \bibinfo{pages}{8527} (\bibinfo{year}{2015}).

\bibitem[{\citenamefont{Glenn et~al.}(2015)\citenamefont{Glenn, Lee, Park,
  Weissleder, Yacoby, Lukin, Lee, Walsworth, and Connolly}}]{Glenn2015}
\bibinfo{author}{\bibfnamefont{D.~R.} \bibnamefont{Glenn}},
  \bibinfo{author}{\bibfnamefont{K.}~\bibnamefont{Lee}},
  \bibinfo{author}{\bibfnamefont{H.}~\bibnamefont{Park}},
  \bibinfo{author}{\bibfnamefont{R.}~\bibnamefont{Weissleder}},
  \bibinfo{author}{\bibfnamefont{A.}~\bibnamefont{Yacoby}},
  \bibinfo{author}{\bibfnamefont{M.~D.} \bibnamefont{Lukin}},
  \bibinfo{author}{\bibfnamefont{H.}~\bibnamefont{Lee}},
  \bibinfo{author}{\bibfnamefont{R.~L.} \bibnamefont{Walsworth}},
  \bibnamefont{and} \bibinfo{author}{\bibfnamefont{C.~B.}
  \bibnamefont{Connolly}}, \bibinfo{journal}{Nat. Methods}
  \textbf{\bibinfo{volume}{12}}, \bibinfo{pages}{736} (\bibinfo{year}{2015}).

\bibitem[{\citenamefont{Unden et~al.}(2016)\citenamefont{Unden,
  Balasubramanian, Louzon, Vinkler, Plenio, Markham, Twitchen, Stacey,
  Lovchinsky, Sushkov et~al.}}]{Unden2016}
\bibinfo{author}{\bibfnamefont{T.}~\bibnamefont{Unden}},
  \bibinfo{author}{\bibfnamefont{P.}~\bibnamefont{Balasubramanian}},
  \bibinfo{author}{\bibfnamefont{D.}~\bibnamefont{Louzon}},
  \bibinfo{author}{\bibfnamefont{Y.}~\bibnamefont{Vinkler}},
  \bibinfo{author}{\bibfnamefont{M.~B.} \bibnamefont{Plenio}},
  \bibinfo{author}{\bibfnamefont{M.}~\bibnamefont{Markham}},
  \bibinfo{author}{\bibfnamefont{D.}~\bibnamefont{Twitchen}},
  \bibinfo{author}{\bibfnamefont{A.}~\bibnamefont{Stacey}},
  \bibinfo{author}{\bibfnamefont{I.}~\bibnamefont{Lovchinsky}},
  \bibinfo{author}{\bibfnamefont{A.~O.} \bibnamefont{Sushkov}},
  \bibnamefont{et~al.}, \bibinfo{journal}{Phys. Rev. Lett.}
  \textbf{\bibinfo{volume}{116}}, \bibinfo{pages}{230502}
  (\bibinfo{year}{2016}).

\bibitem[{\citenamefont{Bonato et~al.}(2016)\citenamefont{Bonato, Blok, Dinani,
  Berry, Markham, Twitchen, and Hanson}}]{Bonato2016}
\bibinfo{author}{\bibfnamefont{C.}~\bibnamefont{Bonato}},
  \bibinfo{author}{\bibfnamefont{M.~S.} \bibnamefont{Blok}},
  \bibinfo{author}{\bibfnamefont{H.~T.} \bibnamefont{Dinani}},
  \bibinfo{author}{\bibfnamefont{D.~W.} \bibnamefont{Berry}},
  \bibinfo{author}{\bibfnamefont{M.~L.} \bibnamefont{Markham}},
  \bibinfo{author}{\bibfnamefont{D.~J.} \bibnamefont{Twitchen}},
  \bibnamefont{and} \bibinfo{author}{\bibfnamefont{R.}~\bibnamefont{Hanson}},
  \bibinfo{journal}{Nat. Nanotechnol.} \textbf{\bibinfo{volume}{11}},
  \bibinfo{pages}{247} (\bibinfo{year}{2016}).

\bibitem[{\citenamefont{Rosi et~al.}(2013)\citenamefont{Rosi, Bernard, Fabbri,
  Fallani, Fort, Inguscio, Calarco, and Montangero}}]{Rosi2013}
\bibinfo{author}{\bibfnamefont{S.}~\bibnamefont{Rosi}},
  \bibinfo{author}{\bibfnamefont{A.}~\bibnamefont{Bernard}},
  \bibinfo{author}{\bibfnamefont{N.}~\bibnamefont{Fabbri}},
  \bibinfo{author}{\bibfnamefont{L.}~\bibnamefont{Fallani}},
  \bibinfo{author}{\bibfnamefont{C.}~\bibnamefont{Fort}},
  \bibinfo{author}{\bibfnamefont{M.}~\bibnamefont{Inguscio}},
  \bibinfo{author}{\bibfnamefont{T.}~\bibnamefont{Calarco}}, \bibnamefont{and}
  \bibinfo{author}{\bibfnamefont{S.}~\bibnamefont{Montangero}},
  \bibinfo{journal}{Phys. Rev. A} \textbf{\bibinfo{volume}{88}},
  \bibinfo{pages}{021601} (\bibinfo{year}{2013}).

\bibitem[{\citenamefont{Brif et~al.}(2010)\citenamefont{Brif, Chakrabarti, and
  Rabitz}}]{Brif2010}
\bibinfo{author}{\bibfnamefont{C.}~\bibnamefont{Brif}},
  \bibinfo{author}{\bibfnamefont{R.}~\bibnamefont{Chakrabarti}},
  \bibnamefont{and} \bibinfo{author}{\bibfnamefont{H.}~\bibnamefont{Rabitz}},
  \bibinfo{journal}{New J. Phys.} \textbf{\bibinfo{volume}{12}},
  \bibinfo{pages}{075008} (\bibinfo{year}{2010}).

\bibitem[{\citenamefont{Doria et~al.}(2011)\citenamefont{Doria, Calarco, and
  Montangero}}]{Doria2011}
\bibinfo{author}{\bibfnamefont{P.}~\bibnamefont{Doria}},
  \bibinfo{author}{\bibfnamefont{T.}~\bibnamefont{Calarco}}, \bibnamefont{and}
  \bibinfo{author}{\bibfnamefont{S.}~\bibnamefont{Montangero}},
  \bibinfo{journal}{Phys. Rev. Lett.} \textbf{\bibinfo{volume}{106}},
  \bibinfo{pages}{190501} (\bibinfo{year}{2011}).

\bibitem[{\citenamefont{Rach et~al.}(2015)\citenamefont{Rach, M\"uller,
  Calarco, and Montangero}}]{Rach2015}
\bibinfo{author}{\bibfnamefont{N.}~\bibnamefont{Rach}},
  \bibinfo{author}{\bibfnamefont{M.~M.} \bibnamefont{M\"uller}},
  \bibinfo{author}{\bibfnamefont{T.}~\bibnamefont{Calarco}}, \bibnamefont{and}
  \bibinfo{author}{\bibfnamefont{S.}~\bibnamefont{Montangero}},
  \bibinfo{journal}{Phys. Rev. A} \textbf{\bibinfo{volume}{92}},
  \bibinfo{pages}{062343} (\bibinfo{year}{2015}).

\bibitem[{\citenamefont{Dolde et~al.}(2014)\citenamefont{Dolde, Bergholm, Wang,
  Jakobi, Naydenov, Pezzagna, Meijer, Jelezko, Neumann, Schulte-Herbr{\"u}ggen
  et~al.}}]{Dolde2014}
\bibinfo{author}{\bibfnamefont{F.}~\bibnamefont{Dolde}},
  \bibinfo{author}{\bibfnamefont{V.}~\bibnamefont{Bergholm}},
  \bibinfo{author}{\bibfnamefont{Y.}~\bibnamefont{Wang}},
  \bibinfo{author}{\bibfnamefont{I.}~\bibnamefont{Jakobi}},
  \bibinfo{author}{\bibfnamefont{B.}~\bibnamefont{Naydenov}},
  \bibinfo{author}{\bibfnamefont{S.}~\bibnamefont{Pezzagna}},
  \bibinfo{author}{\bibfnamefont{J.}~\bibnamefont{Meijer}},
  \bibinfo{author}{\bibfnamefont{F.}~\bibnamefont{Jelezko}},
  \bibinfo{author}{\bibfnamefont{P.}~\bibnamefont{Neumann}},
  \bibinfo{author}{\bibfnamefont{T.}~\bibnamefont{Schulte-Herbr{\"u}ggen}},
  \bibnamefont{et~al.}, \bibinfo{journal}{Nat. Commun.}
  \textbf{\bibinfo{volume}{5}} (\bibinfo{year}{2014}).

\bibitem[{\citenamefont{Waldherr et~al.}(2014)\citenamefont{Waldherr, Wang,
  Zaiser, Jamali, Schulte-Herbruggen, Abe, Ohshima, Isoya, Du, Neumann
  et~al.}}]{Waldherr2014}
\bibinfo{author}{\bibfnamefont{G.}~\bibnamefont{Waldherr}},
  \bibinfo{author}{\bibfnamefont{Y.}~\bibnamefont{Wang}},
  \bibinfo{author}{\bibfnamefont{S.}~\bibnamefont{Zaiser}},
  \bibinfo{author}{\bibfnamefont{M.}~\bibnamefont{Jamali}},
  \bibinfo{author}{\bibfnamefont{T.}~\bibnamefont{Schulte-Herbruggen}},
  \bibinfo{author}{\bibfnamefont{H.}~\bibnamefont{Abe}},
  \bibinfo{author}{\bibfnamefont{T.}~\bibnamefont{Ohshima}},
  \bibinfo{author}{\bibfnamefont{J.}~\bibnamefont{Isoya}},
  \bibinfo{author}{\bibfnamefont{J.~F.} \bibnamefont{Du}},
  \bibinfo{author}{\bibfnamefont{P.}~\bibnamefont{Neumann}},
  \bibnamefont{et~al.}, \bibinfo{journal}{Nature}
  \textbf{\bibinfo{volume}{506}}, \bibinfo{pages}{204} (\bibinfo{year}{2014}).

\bibitem[{\citenamefont{Wang et~al.}(2015)\citenamefont{Wang, Dolde, Biamonte,
  Babbush, Bergholm, Yang, Jakobi, Neumann, Aspuru-Guzik, Whitfield
  et~al.}}]{Wang2015}
\bibinfo{author}{\bibfnamefont{Y.}~\bibnamefont{Wang}},
  \bibinfo{author}{\bibfnamefont{F.}~\bibnamefont{Dolde}},
  \bibinfo{author}{\bibfnamefont{J.}~\bibnamefont{Biamonte}},
  \bibinfo{author}{\bibfnamefont{R.}~\bibnamefont{Babbush}},
  \bibinfo{author}{\bibfnamefont{V.}~\bibnamefont{Bergholm}},
  \bibinfo{author}{\bibfnamefont{S.}~\bibnamefont{Yang}},
  \bibinfo{author}{\bibfnamefont{I.}~\bibnamefont{Jakobi}},
  \bibinfo{author}{\bibfnamefont{P.}~\bibnamefont{Neumann}},
  \bibinfo{author}{\bibfnamefont{A.}~\bibnamefont{Aspuru-Guzik}},
  \bibinfo{author}{\bibfnamefont{J.~D.} \bibnamefont{Whitfield}},
  \bibnamefont{et~al.}, \bibinfo{journal}{ACS Nano}
  \textbf{\bibinfo{volume}{9}}, \bibinfo{pages}{7769} (\bibinfo{year}{2015}).

\bibitem[{\citenamefont{Scheuer et~al.}(2014)\citenamefont{Scheuer, Kong, Said,
  Chen, Kurz, Marseglia, Du, Hemmer, Montangero, Calarco et~al.}}]{Scheuer2014}
\bibinfo{author}{\bibfnamefont{J.}~\bibnamefont{Scheuer}},
  \bibinfo{author}{\bibfnamefont{X.}~\bibnamefont{Kong}},
  \bibinfo{author}{\bibfnamefont{R.~S.} \bibnamefont{Said}},
  \bibinfo{author}{\bibfnamefont{J.}~\bibnamefont{Chen}},
  \bibinfo{author}{\bibfnamefont{A.}~\bibnamefont{Kurz}},
  \bibinfo{author}{\bibfnamefont{L.}~\bibnamefont{Marseglia}},
  \bibinfo{author}{\bibfnamefont{J.}~\bibnamefont{Du}},
  \bibinfo{author}{\bibfnamefont{P.~R.} \bibnamefont{Hemmer}},
  \bibinfo{author}{\bibfnamefont{S.}~\bibnamefont{Montangero}},
  \bibinfo{author}{\bibfnamefont{T.}~\bibnamefont{Calarco}},
  \bibnamefont{et~al.}, \bibinfo{journal}{New J. Phys.}
  \textbf{\bibinfo{volume}{16}}, \bibinfo{pages}{093022}
  (\bibinfo{year}{2014}).

\bibitem[{\citenamefont{N\"obauer et~al.}(2015)\citenamefont{N\"obauer,
  Angerer, Bartels, Trupke, Rotter, Schmiedmayer, Mintert, and
  Majer}}]{Noebauer2015}
\bibinfo{author}{\bibfnamefont{T.}~\bibnamefont{N\"obauer}},
  \bibinfo{author}{\bibfnamefont{A.}~\bibnamefont{Angerer}},
  \bibinfo{author}{\bibfnamefont{B.}~\bibnamefont{Bartels}},
  \bibinfo{author}{\bibfnamefont{M.}~\bibnamefont{Trupke}},
  \bibinfo{author}{\bibfnamefont{S.}~\bibnamefont{Rotter}},
  \bibinfo{author}{\bibfnamefont{J.}~\bibnamefont{Schmiedmayer}},
  \bibinfo{author}{\bibfnamefont{F.}~\bibnamefont{Mintert}}, \bibnamefont{and}
  \bibinfo{author}{\bibfnamefont{J.}~\bibnamefont{Majer}},
  \bibinfo{journal}{Phys. Rev. Lett.} \textbf{\bibinfo{volume}{115}},
  \bibinfo{pages}{190801} (\bibinfo{year}{2015}).

\bibitem[{\citenamefont{Lloyd and Montangero}(2014)}]{Lloyd2014}
\bibinfo{author}{\bibfnamefont{S.}~\bibnamefont{Lloyd}} \bibnamefont{and}
  \bibinfo{author}{\bibfnamefont{S.}~\bibnamefont{Montangero}},
  \bibinfo{journal}{Phys. Rev. Lett.} \textbf{\bibinfo{volume}{113}},
  \bibinfo{pages}{010502} (\bibinfo{year}{2014}).

\bibitem[{\citenamefont{Caneva et~al.}(2014)\citenamefont{Caneva, Silva, Fazio,
  Lloyd, Calarco, and Montangero}}]{Caneva2014}
\bibinfo{author}{\bibfnamefont{T.}~\bibnamefont{Caneva}},
  \bibinfo{author}{\bibfnamefont{A.}~\bibnamefont{Silva}},
  \bibinfo{author}{\bibfnamefont{R.}~\bibnamefont{Fazio}},
  \bibinfo{author}{\bibfnamefont{S.}~\bibnamefont{Lloyd}},
  \bibinfo{author}{\bibfnamefont{T.}~\bibnamefont{Calarco}}, \bibnamefont{and}
  \bibinfo{author}{\bibfnamefont{S.}~\bibnamefont{Montangero}},
  \bibinfo{journal}{Phys. Rev. A} \textbf{\bibinfo{volume}{89}},
  \bibinfo{pages}{042322} (\bibinfo{year}{2014}).

\bibitem[{\citenamefont{Viola and Lloyd}(1998)}]{Viola1998}
\bibinfo{author}{\bibfnamefont{L.}~\bibnamefont{Viola}} \bibnamefont{and}
  \bibinfo{author}{\bibfnamefont{S.}~\bibnamefont{Lloyd}},
  \bibinfo{journal}{Phys. Rev. A} \textbf{\bibinfo{volume}{58}},
  \bibinfo{pages}{2733} (\bibinfo{year}{1998}).

\bibitem[{\citenamefont{O'Hara et~al.}(2001)\citenamefont{O'Hara, Gehm,
  Granade, and Thomas}}]{OHara2001}
\bibinfo{author}{\bibfnamefont{K.~M.} \bibnamefont{O'Hara}},
  \bibinfo{author}{\bibfnamefont{M.~E.} \bibnamefont{Gehm}},
  \bibinfo{author}{\bibfnamefont{S.~R.} \bibnamefont{Granade}},
  \bibnamefont{and} \bibinfo{author}{\bibfnamefont{J.~E.}
  \bibnamefont{Thomas}}, \bibinfo{journal}{Phys. Rev. A}
  \textbf{\bibinfo{volume}{64}}, \bibinfo{pages}{051403}
  (\bibinfo{year}{2001}).

\bibitem[{\citenamefont{Wigley et~al.}(2016)\citenamefont{Wigley, Everitt,
  van~den Hengel, Bastian, Sooriyabandara, McDonald, Hardman, Quinlivan, Manju,
  Kuhn et~al.}}]{Wigley2016}
\bibinfo{author}{\bibfnamefont{P.~B.} \bibnamefont{Wigley}},
  \bibinfo{author}{\bibfnamefont{P.~J.} \bibnamefont{Everitt}},
  \bibinfo{author}{\bibfnamefont{A.}~\bibnamefont{van~den Hengel}},
  \bibinfo{author}{\bibfnamefont{J.~W.} \bibnamefont{Bastian}},
  \bibinfo{author}{\bibfnamefont{M.~A.} \bibnamefont{Sooriyabandara}},
  \bibinfo{author}{\bibfnamefont{G.~D.} \bibnamefont{McDonald}},
  \bibinfo{author}{\bibfnamefont{K.~S.} \bibnamefont{Hardman}},
  \bibinfo{author}{\bibfnamefont{C.~D.} \bibnamefont{Quinlivan}},
  \bibinfo{author}{\bibfnamefont{P.}~\bibnamefont{Manju}},
  \bibinfo{author}{\bibfnamefont{C.~C.~N.} \bibnamefont{Kuhn}},
  \bibnamefont{et~al.}, \bibinfo{journal}{Sci. Rep.}
  \textbf{\bibinfo{volume}{6}}, \bibinfo{pages}{25890} (\bibinfo{year}{2016}).

\bibitem[{\citenamefont{Knill et~al.}(2008)\citenamefont{Knill, Leibfried,
  Reichle, Britton, Blakestad, Jost, Langer, Ozeri, Seidelin, and
  Wineland}}]{Knill2008}
\bibinfo{author}{\bibfnamefont{E.}~\bibnamefont{Knill}},
  \bibinfo{author}{\bibfnamefont{D.}~\bibnamefont{Leibfried}},
  \bibinfo{author}{\bibfnamefont{R.}~\bibnamefont{Reichle}},
  \bibinfo{author}{\bibfnamefont{J.}~\bibnamefont{Britton}},
  \bibinfo{author}{\bibfnamefont{R.~B.} \bibnamefont{Blakestad}},
  \bibinfo{author}{\bibfnamefont{J.~D.} \bibnamefont{Jost}},
  \bibinfo{author}{\bibfnamefont{C.}~\bibnamefont{Langer}},
  \bibinfo{author}{\bibfnamefont{R.}~\bibnamefont{Ozeri}},
  \bibinfo{author}{\bibfnamefont{S.}~\bibnamefont{Seidelin}}, \bibnamefont{and}
  \bibinfo{author}{\bibfnamefont{D.~J.} \bibnamefont{Wineland}},
  \bibinfo{journal}{Phys. Rev. A} \textbf{\bibinfo{volume}{77}},
  \bibinfo{pages}{012307} (\bibinfo{year}{2008}).

\end{thebibliography}

\begin{thebibliography}{23}
\expandafter\ifx\csname natexlab\endcsname\relax\def\natexlab#1{#1}\fi
\expandafter\ifx\csname bibnamefont\endcsname\relax
  \def\bibnamefont#1{#1}\fi
\expandafter\ifx\csname bibfnamefont\endcsname\relax
  \def\bibfnamefont#1{#1}\fi
\expandafter\ifx\csname citenamefont\endcsname\relax
  \def\citenamefont#1{#1}\fi
\expandafter\ifx\csname url\endcsname\relax
  \def\url#1{\texttt{#1}}\fi
\expandafter\ifx\csname urlprefix\endcsname\relax\def\urlprefix{URL }\fi
\providecommand{\bibinfo}[2]{#2}
\providecommand{\eprint}[2][]{\url{#2}}

\bibitem[{\citenamefont{Doria et~al.}(2011)\citenamefont{Doria, Calarco, and
  Montangero}}]{SM_Doria2011}
\bibinfo{author}{\bibfnamefont{P.}~\bibnamefont{Doria}},
  \bibinfo{author}{\bibfnamefont{T.}~\bibnamefont{Calarco}}, \bibnamefont{and}
  \bibinfo{author}{\bibfnamefont{S.}~\bibnamefont{Montangero}},
  \bibinfo{journal}{Phys. Rev. Lett.} \textbf{\bibinfo{volume}{106}},
  \bibinfo{pages}{190501} (\bibinfo{year}{2011}).

\bibitem[{\citenamefont{Caneva et~al.}(2011)\citenamefont{Caneva, Calarco, and
  Montangero}}]{SM_Caneva2011}
\bibinfo{author}{\bibfnamefont{T.}~\bibnamefont{Caneva}},
  \bibinfo{author}{\bibfnamefont{T.}~\bibnamefont{Calarco}}, \bibnamefont{and}
  \bibinfo{author}{\bibfnamefont{S.}~\bibnamefont{Montangero}},
  \bibinfo{journal}{Phys. Rev. A} \textbf{\bibinfo{volume}{84}},
  \bibinfo{pages}{022326} (\bibinfo{year}{2011}).

\bibitem[{\citenamefont{Rach et~al.}(2015)\citenamefont{Rach, M\"uller,
  Calarco, and Montangero}}]{SM_Rach2015}
\bibinfo{author}{\bibfnamefont{N.}~\bibnamefont{Rach}},
  \bibinfo{author}{\bibfnamefont{M.~M.} \bibnamefont{M\"uller}},
  \bibinfo{author}{\bibfnamefont{T.}~\bibnamefont{Calarco}}, \bibnamefont{and}
  \bibinfo{author}{\bibfnamefont{S.}~\bibnamefont{Montangero}},
  \bibinfo{journal}{Phys. Rev. A} \textbf{\bibinfo{volume}{92}},
  \bibinfo{pages}{062343} (\bibinfo{year}{2015}).

\bibitem[{\citenamefont{Scheuer et~al.}(2014)\citenamefont{Scheuer, Kong, Said,
  Chen, Kurz, Marseglia, Du, Hemmer, Montangero, Calarco et~al.}}]{SM_Scheuer2014}
\bibinfo{author}{\bibfnamefont{J.}~\bibnamefont{Scheuer}},
  \bibinfo{author}{\bibfnamefont{X.}~\bibnamefont{Kong}},
  \bibinfo{author}{\bibfnamefont{R.~S.} \bibnamefont{Said}},
  \bibinfo{author}{\bibfnamefont{J.}~\bibnamefont{Chen}},
  \bibinfo{author}{\bibfnamefont{A.}~\bibnamefont{Kurz}},
  \bibinfo{author}{\bibfnamefont{L.}~\bibnamefont{Marseglia}},
  \bibinfo{author}{\bibfnamefont{J.}~\bibnamefont{Du}},
  \bibinfo{author}{\bibfnamefont{P.~R.} \bibnamefont{Hemmer}},
  \bibinfo{author}{\bibfnamefont{S.}~\bibnamefont{Montangero}},
  \bibinfo{author}{\bibfnamefont{T.}~\bibnamefont{Calarco}},
  \bibnamefont{et~al.}, \bibinfo{journal}{New J. Phys.}
  \textbf{\bibinfo{volume}{16}}, \bibinfo{pages}{093022}
  (\bibinfo{year}{2014}).

\bibitem[{\citenamefont{Rosi et~al.}(2013)\citenamefont{Rosi, Bernard, Fabbri,
  Fallani, Fort, Inguscio, Calarco, and Montangero}}]{SM_Rosi2013}
\bibinfo{author}{\bibfnamefont{S.}~\bibnamefont{Rosi}},
  \bibinfo{author}{\bibfnamefont{A.}~\bibnamefont{Bernard}},
  \bibinfo{author}{\bibfnamefont{N.}~\bibnamefont{Fabbri}},
  \bibinfo{author}{\bibfnamefont{L.}~\bibnamefont{Fallani}},
  \bibinfo{author}{\bibfnamefont{C.}~\bibnamefont{Fort}},
  \bibinfo{author}{\bibfnamefont{M.}~\bibnamefont{Inguscio}},
  \bibinfo{author}{\bibfnamefont{T.}~\bibnamefont{Calarco}}, \bibnamefont{and}
  \bibinfo{author}{\bibfnamefont{S.}~\bibnamefont{Montangero}},
  \bibinfo{journal}{Phys. Rev. A} \textbf{\bibinfo{volume}{88}},
  \bibinfo{pages}{021601} (\bibinfo{year}{2013}).

\bibitem[{\citenamefont{van Frank et~al.}(2014)\citenamefont{van Frank,
  Negretti, Berrada, B{\"u}cker, Montangero, Schaff, Schumm, Calarco, and
  Schmiedmayer}}]{SM_Frank2014}
\bibinfo{author}{\bibfnamefont{S.}~\bibnamefont{van Frank}},
  \bibinfo{author}{\bibfnamefont{A.}~\bibnamefont{Negretti}},
  \bibinfo{author}{\bibfnamefont{T.}~\bibnamefont{Berrada}},
  \bibinfo{author}{\bibfnamefont{R.}~\bibnamefont{B{\"u}cker}},
  \bibinfo{author}{\bibfnamefont{S.}~\bibnamefont{Montangero}},
  \bibinfo{author}{\bibfnamefont{J.~F.} \bibnamefont{Schaff}},
  \bibinfo{author}{\bibfnamefont{T.}~\bibnamefont{Schumm}},
  \bibinfo{author}{\bibfnamefont{T.}~\bibnamefont{Calarco}}, \bibnamefont{and}
  \bibinfo{author}{\bibfnamefont{J.}~\bibnamefont{Schmiedmayer}},
  \bibinfo{journal}{Nat Commun} \textbf{\bibinfo{volume}{5}}
  (\bibinfo{year}{2014}).

\bibitem[{\citenamefont{Brouzos et~al.}(2015)\citenamefont{Brouzos, Streltsov,
  Negretti, Said, Caneva, Montangero, and Calarco}}]{SM_Brouzos2015}
\bibinfo{author}{\bibfnamefont{I.}~\bibnamefont{Brouzos}},
  \bibinfo{author}{\bibfnamefont{A.~I.} \bibnamefont{Streltsov}},
  \bibinfo{author}{\bibfnamefont{A.}~\bibnamefont{Negretti}},
  \bibinfo{author}{\bibfnamefont{R.~S.} \bibnamefont{Said}},
  \bibinfo{author}{\bibfnamefont{T.}~\bibnamefont{Caneva}},
  \bibinfo{author}{\bibfnamefont{S.}~\bibnamefont{Montangero}},
  \bibnamefont{and} \bibinfo{author}{\bibfnamefont{T.}~\bibnamefont{Calarco}},
  \bibinfo{journal}{Phys. Rev. A} \textbf{\bibinfo{volume}{92}},
  \bibinfo{pages}{062110} (\bibinfo{year}{2015}).

\bibitem[{\citenamefont{Alon et~al.}(2016)\citenamefont{Alon, Bagnato, Beinke,
  Brouzos, Calarco, Caneva, Cederbaum, Kasevich, Klaiman, Lode
  et~al.}}]{SM_Alon2016}
\bibinfo{author}{\bibfnamefont{O.~E.} \bibnamefont{Alon}},
  \bibinfo{author}{\bibfnamefont{V.~S.} \bibnamefont{Bagnato}},
  \bibinfo{author}{\bibfnamefont{R.}~\bibnamefont{Beinke}},
  \bibinfo{author}{\bibfnamefont{I.}~\bibnamefont{Brouzos}},
  \bibinfo{author}{\bibfnamefont{T.}~\bibnamefont{Calarco}},
  \bibinfo{author}{\bibfnamefont{T.}~\bibnamefont{Caneva}},
  \bibinfo{author}{\bibfnamefont{L.~S.} \bibnamefont{Cederbaum}},
  \bibinfo{author}{\bibfnamefont{M.~A.} \bibnamefont{Kasevich}},
  \bibinfo{author}{\bibfnamefont{S.}~\bibnamefont{Klaiman}},
  \bibinfo{author}{\bibfnamefont{A.~U.~J.} \bibnamefont{Lode}},
  \bibnamefont{et~al.}, \emph{\bibinfo{title}{High Performance Computing in
  Science and Engineering 15: Transactions of the High Performance Computing
  Center, Stuttgart (HLRS) 2015}} (\bibinfo{publisher}{Springer International
  Publishing}, \bibinfo{address}{Cham}, \bibinfo{year}{2016}), chap.
  \bibinfo{chapter}{MCTDHB Physics and Technologies: Excitations and Vorticity,
  Single-Shot Detection, Measurement of Fragmentation, and Optimal Control in
  Correlated Ultra-Cold Bosonic Many-Body Systems}, pp.
  \bibinfo{pages}{23--49}.

\bibitem[{\citenamefont{van Frank et~al.}(2016)\citenamefont{van Frank,
  Bonneau, Schmiedmayer, Hild, Gross, Cheneau, Bloch, Pichler, Negretti,
  Calarco et~al.}}]{SM_Frank2016}
\bibinfo{author}{\bibfnamefont{S.}~\bibnamefont{van Frank}},
  \bibinfo{author}{\bibfnamefont{M.}~\bibnamefont{Bonneau}},
  \bibinfo{author}{\bibfnamefont{J.}~\bibnamefont{Schmiedmayer}},
  \bibinfo{author}{\bibfnamefont{S.}~\bibnamefont{Hild}},
  \bibinfo{author}{\bibfnamefont{C.}~\bibnamefont{Gross}},
  \bibinfo{author}{\bibfnamefont{M.}~\bibnamefont{Cheneau}},
  \bibinfo{author}{\bibfnamefont{I.}~\bibnamefont{Bloch}},
  \bibinfo{author}{\bibfnamefont{T.}~\bibnamefont{Pichler}},
  \bibinfo{author}{\bibfnamefont{A.}~\bibnamefont{Negretti}},
  \bibinfo{author}{\bibfnamefont{T.}~\bibnamefont{Calarco}},
  \bibnamefont{et~al.}, \bibinfo{journal}{Scientific Reports}
  \textbf{\bibinfo{volume}{6}}, \bibinfo{pages}{34187 EP }
  (\bibinfo{year}{2016}).

\bibitem[{\citenamefont{Caneva et~al.}(2009)\citenamefont{Caneva, Murphy,
  Calarco, Fazio, Montangero, Giovannetti, and Santoro}}]{SM_Caneva2009}
\bibinfo{author}{\bibfnamefont{T.}~\bibnamefont{Caneva}},
  \bibinfo{author}{\bibfnamefont{M.}~\bibnamefont{Murphy}},
  \bibinfo{author}{\bibfnamefont{T.}~\bibnamefont{Calarco}},
  \bibinfo{author}{\bibfnamefont{R.}~\bibnamefont{Fazio}},
  \bibinfo{author}{\bibfnamefont{S.}~\bibnamefont{Montangero}},
  \bibinfo{author}{\bibfnamefont{V.}~\bibnamefont{Giovannetti}},
  \bibnamefont{and} \bibinfo{author}{\bibfnamefont{G.~E.}
  \bibnamefont{Santoro}}, \bibinfo{journal}{Phys. Rev. Lett.}
  \textbf{\bibinfo{volume}{103}}, \bibinfo{pages}{240501}
  (\bibinfo{year}{2009}).

\bibitem[{\citenamefont{Nelder and Mead}(1965)}]{SM_Nelder1965}
\bibinfo{author}{\bibfnamefont{J.~A.} \bibnamefont{Nelder}} \bibnamefont{and}
  \bibinfo{author}{\bibfnamefont{R.}~\bibnamefont{Mead}}, \bibinfo{journal}{The
  Computer Journal} \textbf{\bibinfo{volume}{7}}, \bibinfo{pages}{308}
  (\bibinfo{year}{1965}).

\bibitem[{\citenamefont{Powell}(1964)}]{SM_Powell1964}
\bibinfo{author}{\bibfnamefont{M.~J.~D.} \bibnamefont{Powell}},
  \bibinfo{journal}{The Computer Journal} \textbf{\bibinfo{volume}{7}},
  \bibinfo{pages}{155} (\bibinfo{year}{1964}).

\bibitem[{\citenamefont{Audet and J.~E.~Dennis}(2002)}]{SM_Audet2002}
\bibinfo{author}{\bibfnamefont{C.}~\bibnamefont{Audet}} \bibnamefont{and}
  \bibinfo{author}{\bibfnamefont{J.}~\bibnamefont{J.~E.~Dennis}},
  \bibinfo{journal}{SIAM Journal on Optimization}
  \textbf{\bibinfo{volume}{13}}, \bibinfo{pages}{889} (\bibinfo{year}{2002}).

\bibitem[{\citenamefont{Lagarias et~al.}(1998)\citenamefont{Lagarias, Reeds,
  Wright, and Wright}}]{SM_Lagarias1998}
\bibinfo{author}{\bibfnamefont{J.~C.} \bibnamefont{Lagarias}},
  \bibinfo{author}{\bibfnamefont{J.~A.} \bibnamefont{Reeds}},
  \bibinfo{author}{\bibfnamefont{M.~H.} \bibnamefont{Wright}},
  \bibnamefont{and} \bibinfo{author}{\bibfnamefont{P.~E.}
  \bibnamefont{Wright}}, \bibinfo{journal}{SIAM Journal on Optimization}
  \textbf{\bibinfo{volume}{9}}, \bibinfo{pages}{112} (\bibinfo{year}{1998}).

\bibitem[{\citenamefont{Lewis et~al.}(2000)\citenamefont{Lewis, Torczon, and
  Trosset}}]{SM_Lewis2000}
\bibinfo{author}{\bibfnamefont{R.~M.} \bibnamefont{Lewis}},
  \bibinfo{author}{\bibfnamefont{V.}~\bibnamefont{Torczon}}, \bibnamefont{and}
  \bibinfo{author}{\bibfnamefont{M.~W.} \bibnamefont{Trosset}},
  \bibinfo{journal}{Journal of Computational and Applied Mathematics}
  \textbf{\bibinfo{volume}{124}}, \bibinfo{pages}{191 } (\bibinfo{year}{2000}).

\bibitem[{opt()}]{SM_optimize}
\urlprefix\url{http://docs.scipy.org/doc/scipy/reference/tutorial/optimize.html}.

\bibitem[{fmi()}]{SM_fminsearch}
\urlprefix\url{http://de.mathworks.com/help/matlab/ref/fminsearch.html}.

\bibitem[{\citenamefont{Frank}(2015)}]{SM_Frank2015}
\bibinfo{author}{\bibfnamefont{F.}~\bibnamefont{Frank}}, Master's thesis,
  \bibinfo{school}{University of Ulm} (\bibinfo{year}{2015}).

\bibitem[{\citenamefont{Bayer et~al.}(2008)\citenamefont{Bayer, Wollenhaupt,
  and Baumert}}]{SM_Bayer2008}
\bibinfo{author}{\bibfnamefont{T.}~\bibnamefont{Bayer}},
  \bibinfo{author}{\bibfnamefont{M.}~\bibnamefont{Wollenhaupt}},
  \bibnamefont{and} \bibinfo{author}{\bibfnamefont{T.}~\bibnamefont{Baumert}},
  \bibinfo{journal}{Journal of Physics B: Atomic, Molecular and Optical
  Physics} \textbf{\bibinfo{volume}{41}}, \bibinfo{pages}{074007}
  (\bibinfo{year}{2008}).

\bibitem[{\citenamefont{Roslund and Rabitz}(2009{\natexlab{a}})}]{SM_Roslund2009}
\bibinfo{author}{\bibfnamefont{J.}~\bibnamefont{Roslund}} \bibnamefont{and}
  \bibinfo{author}{\bibfnamefont{H.}~\bibnamefont{Rabitz}},
  \bibinfo{journal}{Phys. Rev. A} \textbf{\bibinfo{volume}{79}},
  \bibinfo{pages}{053417} (\bibinfo{year}{2009}{\natexlab{a}}).

\bibitem[{\citenamefont{Roslund and Rabitz}(2009{\natexlab{b}})}]{SM_Roslund2009a}
\bibinfo{author}{\bibfnamefont{J.}~\bibnamefont{Roslund}} \bibnamefont{and}
  \bibinfo{author}{\bibfnamefont{H.}~\bibnamefont{Rabitz}},
  \bibinfo{journal}{Phys. Rev. A} \textbf{\bibinfo{volume}{80}},
  \bibinfo{pages}{013408} (\bibinfo{year}{2009}{\natexlab{b}}).

\bibitem[{\citenamefont{Neumann et~al.}(2008)\citenamefont{Neumann, Mizuochi,
  Rempp, Hemmer, Watanabe, Yamasaki, Jacques, Gaebel, Jelezko, and
  Wrachtrup}}]{SM_Neumann1326}
\bibinfo{author}{\bibfnamefont{P.}~\bibnamefont{Neumann}},
  \bibinfo{author}{\bibfnamefont{N.}~\bibnamefont{Mizuochi}},
  \bibinfo{author}{\bibfnamefont{F.}~\bibnamefont{Rempp}},
  \bibinfo{author}{\bibfnamefont{P.}~\bibnamefont{Hemmer}},
  \bibinfo{author}{\bibfnamefont{H.}~\bibnamefont{Watanabe}},
  \bibinfo{author}{\bibfnamefont{S.}~\bibnamefont{Yamasaki}},
  \bibinfo{author}{\bibfnamefont{V.}~\bibnamefont{Jacques}},
  \bibinfo{author}{\bibfnamefont{T.}~\bibnamefont{Gaebel}},
  \bibinfo{author}{\bibfnamefont{F.}~\bibnamefont{Jelezko}}, \bibnamefont{and}
  \bibinfo{author}{\bibfnamefont{J.}~\bibnamefont{Wrachtrup}},
  \bibinfo{journal}{Science} \textbf{\bibinfo{volume}{320}},
  \bibinfo{pages}{1326} (\bibinfo{year}{2008}).

\bibitem[{\citenamefont{Nielsen and Chuang}(2011)}]{SM_Nielsen2011}
\bibinfo{author}{\bibfnamefont{M.~A.} \bibnamefont{Nielsen}} \bibnamefont{and}
  \bibinfo{author}{\bibfnamefont{I.~L.} \bibnamefont{Chuang}},
  \emph{\bibinfo{title}{Quantum Computation and Quantum Information: 10th
  Anniversary Edition}} (\bibinfo{publisher}{Cambridge University Press},
  \bibinfo{address}{New York, NY, USA}, \bibinfo{year}{2011}).

\end{thebibliography}
\end{document}